\documentclass[twoside]{article}
\usepackage[a4paper]{geometry}
\usepackage[utf8]{inputenc}
\usepackage[T1]{fontenc}
\usepackage{RR}
\usepackage{aeguill} 
\usepackage{amsfonts,amsmath,amssymb} 
\usepackage{eurosym} 
\usepackage[babel]{microtype} 
\usepackage{graphicx}
\usepackage[numbers,square,sort&compress]{natbib}
\usepackage[defblank]{paralist} 
\usepackage{parskip} \parskip=1ex plus 5pt minus 1pt \parindent=4ex 
\usepackage[obeyspaces]{url} 
\usepackage[verbatim]{svn-multi}
\usepackage{xcolor}
\usepackage{xspace}
\usepackage{hyperref}  
    \newcommand{\hhref}[2]{\href{#2}{#1}}

\newcommand{\code}[1]{{\sf #1}}

\newcommand{\Aspirin}{\code{Aspirin}}

\RRNo{9145}
\RRdate{January 2018}
\RRauthor{
Marc Shapiro%
    \thanks{Sorbonne-Université---LIP6---Inria Paris}
  \and
Annette Bieniusa%
\thanks{T.\ U.\ Kaiserslautern}
\and
Nuno Preguiça%
    \thanks{U.\ Nova de Lisboa}
\and
Valter Balegas%
    \thanks{U.\ Nova de Lisboa}
\and
Christopher Meiklejohn%
    \thanks{U.\ Catholique de Louvain}
}
\authorhead{Shapiro \emph{et al.}}
\RRtitle{La juste cohérence pour reconcilier disponibilité et sûreté}
\RRetitle{Just-Right Consistency: reconciling availability and safety}
\RRresume{
  Le théorème CAP, un système de stockage réparti peut être, en cas de
  partition, soit cohérent (CP), soit disponible (AP), mais pas les deux.
  Il y a donc des bases de données CP, où les mises à jour sont
  synchrones, et les bases AP, où elles sont asynchrones.
  Cependant, il n'y a pas de raison essentielle de traiter toutes les
  mises à jour de façon identique.
  L'objectif est que le système reste aussi disponible que possible,
  mais suffisamment synchronisé pour que l'application reste correcte.
  Nous proposons un nouveau principe, la \emph{juste cohérence}, afin de
  concevoir de telles applications, réconciliant la sûreté avec la
  disponibilité et l'efficacité, à partir des constatations suivantes~:
 \begin{inparaenum}[\it (i)]
  \item
    Le modèle de données des  CRDT (Conflict-Free Replicated Data Type)
    permet les mises à jour concurrentes de façon à la fois
    théoriquement fondée et intuitive.
  \item
    Les invariants basés sur la simultanéité ou l'ordre partiel des
    mises à jour sont compatibles avec AP, et peuvent être garanties par
    la Cohérence Causale Transactionelle (TCC), le modèle de cohérence
    le plus fort qui ne compromet pas la disponibilité.

    En ce qui concerne les autres invariants, dits \emph{CAP-sensibles}~:
  \item
    Le cas courant du compteur borné peut être géré par un type de
    données encapsulé, correct et cohérent, appelé \emph{Bounded Counter}~;
    et
  \item
    dans le cas général, une analyse statique permet d'indentifier les
    cas où la sûreté ne nécessite pas de synchronisation.
  \end{inparaenum}
  Notre base de données ``nuage'' Antidote offre les CRDT, le modèle
  TCC, et le type de données Bounded Counter.
  Des outils d'analyse statique et de preuve des invariants
  CAP-sensibles aident à la conception des applications.
  Notre système est mûr pour des applications d'échelle industrielle, et
  a été testé expérimentallement sur des centaines de serveurs répartis
  entre plusieurs centres de données géo-distribués.
}

\RRabstract{
  By the CAP Theorem, a distributed data storage system can ensure
  either Consistency under Partition (CP) or Availability under Partition
  (AP), but not both.
  This has led to a split between CP databases, in which updates are
  synchronous, and AP databases, where they are asynchronous.
  However, there is no inherent reason to treat all updates identically:
  simply, the system should be as available as possible, and synchronised
  just enough for the application to be correct.
  We offer a principled \emph{Just-Right Consistency} approach to
  designing such applications, reconciling correctness with availability
  and performance, based on the following insights:
 \begin{inparaenum}[\it (i)]
  \item
    The Conflict-Free Replicated Data Type (CRDTs) data model supports
    asynchronous updates in an intuitive and principled way.
  \item
    Invariants involving joint or mutually-ordered updates
    are compatible with AP and can be guaranteed by Transactional Causal
    Consistency, the strongest consistency model that does not
    compromise availability.
    Regarding the remaining, ``CAP-sensitive'' invariants:
  \item
    For the common pattern of Bounded Counters, we provide
    encapsulated data type that is proven correct and is efficient;
    and
  \item
    in the general case, static analysis can identify when synchronisation
    is not necessary for correctness.
 \end{inparaenum}
 Our Antidote cloud database system supports CRDTs, Transactional Causal
 Consistency and the Bounded Counter data type.
 Support tools help design applications by static analysis and proof of
 CAP-sensitive invariants.
 This system supports industrial-grade applications and has been tested
 experimentally with hundreds of servers across several geo-distributed
 data centres.
}

\RRmotcle{
  Système distribué~;
  programmation répartie~;
  cohérence~;
  disponibilité~;
  invariants~;
  théorème CAP
}
\RRkeyword{
  Distributed systems;
  distributed programming;
  consistency;
  availability;
  invariants;
  CAP Theorem
}

\RRprojet{Delys}
\URRocq 

\begin{document}
\makeRR   
\tableofcontents
\cleardoublepage{}

\section{Introduction: The CAP gap}
\label{sec:intro}

Many modern applications store their data in a \emph{cloud database
  system} in order to distribute and replicate data over many servers,
across geo-distributed data centres (DCs).
Application developers are faced with unfamiliar behaviours and
unpredictability, bringing complexity and new opportunities for error.

Therefore, some database systems provide ``strong consistency,''
which mimics a sequential, centralised system.
Examples include the Spanner{\slash}F1 system \cite{rep:pan:1693} or the
Serialisability model.%
\footnote{
  Serialisability is characterised by the ACID properties:
  All-or-Nothing, Correct-Individually, Isolation or total order, and
  Durability.
  The first three are defined later in this paper.
  Durability means that the result of some update is observed by all
  later reads, for some definition of ``later.''
}
Under the hood, operations synchronise to maintain the illusion of a
total order.
If the network is \emph{partitioned}, synchronisation blocks and the
database waits indefinitely (until the partition is repaired): the system
remains \emph{Consistent under Partition (CP)}, but is not available.
As synchronisation between geographically remote DCs waits for tens or
hundreds of milliseconds, this has a performance cost.
Thus, Spanner{\slash}F1 requires around 100\,ms to commit an update
transaction \cite{rep:pan:1693}.
CP is overly conservative for many applications.

Alternatively, the system might access the local replica without
synchronising.
Latency is minimal, transactions run in parallel, and the system is
\emph{Available under Partition (AP)} \cite{rep:1691}.
However, a read might return a stale value and writes may conflict.
In this vein, early AP systems, such as Cassandra \cite{db:rep:1768},
Dynamo \cite{app:rep:optim:1606} or Riak \cite{riak} implement Eventual
Consistency, which provides very weak guarantees.

The CAP Theorem states that a system cannot be both CP and AP
\cite{rep:pan:1628}.
%
It seems there are only two alternatives: a conservative CP system that
makes guarantees, but has high cost and low availability; or a bold AP
system that is efficient and available, but has correctness problems.

This is a false dichotomy.
Our insight is to \emph{tailor consistency to application requirements},
and not shoe-horn applications into a rigidly-defined consistency model.
We consider the system is correct if it maintains application-level
\emph{integrity invariants}.
Different applications (or even parts thereof) have very different
invariants.
For instance, many social networks work fine above an AP database.
In contrast, a banking application would seem to require CP to maintain
the ``no overdraft'' invariant.
Notice however that this application is still correct if \code{deposit}
operations are non-synchronised (i.e., AP) \cite{rep:syn:1690};
\code{withdrawal}s themselves must synchronise only when
the balance is low \cite{extending-numeric-2015-09}!

In the rest of this paper, we build upon this intuition to develop the
\emph{Just-Right Consistency} approach.
Our aims is to make the application as available as possible, but
synchronised enough to remain correct.
Our base model, Transactional Causal Consistency, maintains
\emph{AP-compatible} invariant patterns, and we switch to CP selectively
when provably required for a \emph{CAP-sensitive} invariant.

\section{Keep my app safe!}
\label{sec:appl-invar}

To be concrete, let's consider an example, the FMKe application
\cite{app:rep:1774}.
FMKe is modelled after the Danish National Joint Medicine Card system
FMK (\emph{Fælles Medicinkort}), which concerns every Danish
citizen and has been running $24\times 7$ since 2013 \cite{app:rep:1774}.
FMKe handles the lifecycle of prescriptions and events associated
with patients, doctors, hospitals and pharmacies.
Its major operations are the following (each application-level operation
comprises a number of database reads and{\slash}or updates):
\begin{compactitem}
\item \code{create-prescription} creates a prescription record
  associated with a patient, a doctor, and a pharmacy.
\item \code{update-prescription-medication} adds a medication to a
  prescription, or
  increases the \code{count} associated with a medication.
\item \code{process-prescription} corresponds to delivering medication
  by a pharmacy.
\item \code{get-staff-prescriptions} and
  \code{get-pharmacy-prescriptions} return the prescriptions 
  associated with a given staff member and pharmacy respectively.
\end{compactitem}


Let's first consider how this application maintains its invariants in a
sequential setting, in order to identify important patterns.



\paragraph{Relative-order pattern}

Our first programming pattern for preserving invariants leverages the
relative order of database accesses.
%
For instance, in FMKe, \code{create-prescription} first initialises a
prescription record, then makes the relevant patient record point to it.
Changing this order would violate the ``referential integrity''
invariant.





\paragraph{Joint-update pattern}

Our second pattern concerns joint
updates to separate data items.
%
FMKe offers several examples.
For instance, creating a prescription updates not just the patient
record but also the corresponding doctor and pharmacy records.
When the underlying database is non-normalised, FMKe maintains separate
but identical copies of the prescription in each of these records.
In both cases, any other operation accessing the database must observe
the state, either before the joint update takes place (there was no
change), or after (all the joint changes took effect), and will never
observe an intermediate state.



\paragraph{Precondition-check pattern}

The final pattern is conditioning an update to a \emph{precondition check}.
%
For instance, a prescribed medication comes with a \code{count} of how
many times it can be delivered to the patient.
To deliver one box of the medication, \code{process-prescription}
checks {precondition} $\code{count} \ge 1$, and, if true,
decrements \code{count} by one.



\paragraph{The Correct-Individually assumption}



Even when the developer does not think explicitly in terms of
invariants, she uses the above patterns, individually or in combination,
to maintain invariants implicitly.
Informally, relative-order updates maintain a partial order between data
items; joint-updates maintain equivalence between different instances of
the same information; and precondition checks serve to maintain
value-based assertions \cite{syn:formel:sh190}.
The developer must be careful to apply these patterns to maintain the
underlying invariants, even in sequential code.%
\footnote{
  We argue that these three patterns are the critical ones.
  Indirect evidence for this conjecture is that the strongest
  consistency models preserve these patterns, and transparently guarantee
  application invariants \cite{syn:formel:sh190}.
}

Thus, we can make the critical assumption that the application ``does
the right thing'' in a sequential execution (if it is
incorrect sequentially, then discussion of consistency is moot).
Technically, we require that, if the invariants are true in some state
of the database and an operation executes, in the state after the
operation the invariants remain true.
We say that each operation is \emph{correct individually} (the ``C'' in
ACID).

\section{AP-compatible invariant patterns}
\label{sec:AP-compatible}



Reasoning about asynchronous updates is difficult.
This section first presents the CRDT data model; then we discuss an AP
consistency model that preserves the relative-order and joint-update
invariant patterns.

\subsection{Data model: CRDTs}
\label{sec:data-model:-crdts}

Because concurrent assignments do not commute, they may not be
concurrent: ordinary assignments require the CP sledgehammer.
AP data needs an alternative data model suitable for concurrent updates.

A common approach (e.g., in Cassandra) is ``Last-Writer-Wins,'' where
concurrent assignments to the same location are resolved in favour of
the one with the highest timestamp; the other one is a ``lost update.''
A higher-level approach uses Conflict-free Replicated Data Types
(CRDTs) \cite{syn:rep:sh143}.
A CRDT extends a sequential abstract data type, and ensures by
construction that concurrent updates are merged deterministically and
replicas converge.
There are CRDTs for many familiar abstractions, including registers,
counters, sets, maps, graphs and sequences.
For instance, two doctors could concurrently add elements \code{Aspirin}
and \code{Chamomile} to a CRDT set object; as expected, both elements
will be in the set.
Non-commuting updates are resolved according to application
requirements; for instance, when concurrently adding and removing the
same element, \code{add} might win.
Due to space constraints, we refer the reader interested in knowing more
about CRDTs to the literature.

\subsection{The relative-order pattern: Causal Consistency}
\label{sec:PO}

Remember how ensuring the referential integrity invariant relies on the
order in which operations occur.
More generally, applications often use ordering to ensure an
\emph{implication} invariant $P \implies Q$, by first making $Q$ true,
then $P$.%
%
\footnote{
  This approach is derived from the application-level ``demarcation
  protocol'' \cite{syn:1735}.
}
If some replica observed the updates in a different order, the
invariant could be violated.

%
Here is an example.
Database BuggyDB comes with a default admin password of
\code{0000} and admin login disabled.
Cindy, working from the Copenhagen replica, sets the password to
\code{S3kr3t}, then enables login.
BuggyDB does not guarantee to make these updates visible in the same
order.
Malicious user Moriarty, who is accessing a replica in Middelfart
(centre of Denmark), notices admin login is enabled and gains
control with the default password \code{0000}.
The (implicit) invariant ``admin login enabled
$\implies$ password $\neq$ default'' has been violated.

To transparently guarantee the relative-order pattern, the system should
ensure that related updates become visible in the same order at all
replicas.
%
A common approach, called Causal Consistency (CC)
\cite{syn:mat:1025}, is based on Lamport's
happened-before relation \cite{con:rep:615}:
\begin{inparaenum}[\it (i)]
\item \label{program-order}
  If an application thread performs update $u$ followed by update $v$, or
\item \label{write-read}
  if the application reads from update $u$ and later performs
  update $v$, or
\item \label{transitive}
  any transitive combination of the above,  
\end{inparaenum}
then a CC database makes $u$ visible before $v$.
Unrelated ({concurrent}) updates can become visible in
any order.
CC does not impact availability, because the database can always read
some version, and concurrent CRDT writes are merged.

If we assume that the application performs its updates in the right order
(which it must do, since otherwise even sequential execution would be
incorrect), Causal Consistency guarantees that the corresponding
invariants will be maintained transparently.
No extra work is required from the developer; in particular, she does
not need to explicitly understand or write out the invariants.

\subsection{Joint-update pattern: AP transactions}
\label{sec:EQ}

A joint update is a limited form of transaction \cite{db:syn:1751}.
It requires the \emph{All-or-Nothing} property (the ``A'' in ACID).
An FMKe example is creating a prescription, which jointly updates the
corresponding patient, doctor and pharmacy records.

One part of All-or-Nothing is to ensure that, at every replica, either
all the updates of a transaction are visible together, or none is at all.


An incorrectly designed system might violate this property.
Example:
Dr.~Alice in Aalborg (a city in the North of Denmark) adds \Aspirin{} to
Bob's prescription.
This updates both Bob's patient copy and the pharmacy copy.
The underlying BuggyDB2 database in Aalborg pushes its patient-record
updates to the replica in Byrum (on the Læsø island), but not its
pharmacy-record updates (perhaps they are assigned to different servers).
Bob's local pharmacy in Byrum observes that \Aspirin{} appears in Bob's
prescription but, incorrectly, not in the pharmacy's copy.

Clearly, the system should instead transport all the updates of a
transaction as a single unit, even if they belong to different servers
(this property is called atomic writes).
This does not impact availability: if Byrum is partitioned from Aalborg,
Byrum sees no change; after communication is restored, Byrum sees both
updates.
Either way, both sites remain available to their local users.


The complementary \emph{snapshot property} is often overlooked.
It states that all of a transaction's reads must come from (updates by)
the same set of transactions, called its snapshot, even if 
reads are served by different replicas.

%
Let's consider the previous example above BuggyDB3, which implements
atomic writes but not snapshots.
Suppose that Dr.~Alice created Bob's prescription in Transaction T1,
then added \Aspirin{} in Transaction T2.
Transaction T3 reads the patient record written by T1
and the pharmacy record written by T2.
It will find \Aspirin{} in the latter's copy of the prescription but
not in the former, violating their equality.
%

All-or-Nothing (the conjunction of atomic writes and snapshots)
avoids such ``broken reads;'' snapshots are also instrumental in
ensuring that transactions satisfy Causal Consistency.
Let's say T3's snapshot contains T1 and T2.
Then T3 would observe the prescription  set by T1, and both
prescription updates of T2.

If the application developer carefully groups its operations into
transactions (a small price to pay), a database that ensures the
All-or-Nothing properties will transparently guarantee the
corresponding invariants.
No extra work is required from the developer; in particular, she does
not need to explicitly understand or write out the 
invariants.

\subsection{Transactional Causal Consistency, the strongest AP model}
\label{sec:TCC}

Transactions and Causal Consistency are found in many CP models, such as
Strict Serialisability or Snapshot Isolation.
However, many database models (even CP ones, such as Serialisability) do
not enforce condition \emph{(\ref{program-order})} of Causal
Consistency, and therefore could fail the password example.

First-generation AP systems, such as Cassandra \cite{db:rep:1768} or
Riak \cite{riak}, do not support transactions nor Causal Consistency.
This is unfortunate, because these mechanisms are compatible with AP\@.
Without them, developers have a hard time to reason about the behaviour
of their application, as some basic
expectations are violated.
Many applications have implicit invariants that require proper ordering
and grouping \cite{syn:1749}.


The AP model that enforces Causal Consistency and All-or-Nothing is
called Transactional Causal Consistency (TCC).
TCC is the strongest model that does not compromise availability.%
\footnote{
  \citet{syn:rep:1738} call Causal Consistency the strongest AP model,
  but they do not consider transactions, only single operations.
}
Recent research systems such as COPS \cite{rep:syn:1662},
Eiger \cite{syn:rep:1708}, GentleRain \cite{db:syn:1752} or SwiftCloud
\cite{rep:pan:sh177} provide restricted variants of TCC\@.
The open-source, CRDT-based database Antidote, which we developed in the
SyncFree European Project, is the first industrial-strength (now in alpha)
geo-replicated database system with a fully functional and unrestricted
implementation of TCC \cite{rep:pro:sh182, antidote-website}.

\section{CAP-sensitive invariant patterns}
\label{sec:total-order}


Finally, we discuss invariants that are \emph{not} AP-compatible.
Before we discuss the general case, in
Section~\ref{sec:pb-precondition-checks}, let's first consider how to
address a restricted but useful case.

\subsection{A specific case: Bounded Counter data type}
\label{sec:bounded-counter-crdt}

%

A common case of CAP-sensitive problem is maintaining a shared
counter $x$, which supports \code{increment} and \code{decrement}
operations but must remain above some parameter $k$.
By applying \emph{escrow} techniques \cite{db:syn:1689}, carefully
caching partial state, batching synchronisation,
and moving communication off the critical path, the counter can maintain
the invariant $x \ge k$, while remaining efficient and
mostly-AP.

We implement this algorithm in a specialised data type, the Bounded
Counter \cite{extending-numeric-2015-09}.
Skipping the technical details, we illustrate Bounded Counter with an
example.
%
Consider maintaining the budget of the health system with a Bounded
Counter constrained to remain non-negative ($k = 0$).
It may be \code{increment}ed (e.g., receiving payments) or
\code{decrement}ed (e.g., purchasing inventory).
Clearly, \code{increment}s cannot violate the invariant; 
therefore \code{increment} can run in AP mode.
Furthermore, a pharmacy might \code{donate} some of its available share
to another one, even before it is needed; \code{donate} is also an AP
operation.%
\footnote{
  Operation \code{donate} uses the demarcation protocol \cite{syn:1735},
  which requires Causal Consistency.
}
This is
in contrast to \code{decrement};
however, instead of synchronising every \code{decrement}, we can
pre-allocate some share of the budget to each pharmacy and hospital.
As long as the local share remains sufficient,
\code{decrement} affects only the local share, in AP mode.
It is only if the local share is too small that \code{decrement} must
synchronise.
Note that \code{decrement} risks unavailability only in this rare
case.
%

We encapsulate the implementation within the Bounded Counter provided
with Antidote.
The developer does not need to understand the details; she just needs to
set the bound, initial value and initial per-replica share.
Then the application calls \code{increment}, \code{decrement}, and possibly
\code{donate}, as appropriate for the application; synchronisation
remains transparent.
The Bounded Counter algorithm has been proven correct using the general
techniques of Section~\ref{sec:CISE-tool}.


\subsection{The problem with precondition checks}
\label{sec:pb-precondition-checks}

Let's now (finally!) consider the general case of the precondition-check
pattern, of which Bounded Counter is just a restricted example.
%
Unfortunately, this pattern is \emph{CAP-sensitive}, because checking
the local replica might be unsafe in an AP system: even if two replicas
have the same state, one might test the precondition to be true, while
concurrently the other replica is making an update that causes it to
become false; when the second update gets delivered to the first
replica, the invariant is violated.
We say the precondition is not \emph{stable under concurrent update}
\cite{syn:app:sh179}.

%
In FMKe, \code{process-prescription} checks that $\code{count} \ge 1$
and decrements \code{count}, in order to avoid that medication is
delivered in duplicate.
Now, let's say Bob in Byrum has a prescription for one box of \code{Aspirin}.
Bob, and his accomplice Moriarty in Middelfart, present this same
prescription to their local pharmacies at the same time.
At both replicas, the precondition $\code{count} \ge 1$ holds;
inherently to AP, a pharmacy is unaware of the other's concurrent
actions; both decrement \code{count} and deliver the medication,
incorrectly.
The reason is that the precondition evaluates to true at the first
replica, but is negated by the concurrent execution of
\code{process-prescription} at another.
%
%


The only way to be sure the invariant will not be violated is to
prohibit this concurrency (i.e., to synchronise).
Must we admit defeat, and adopt the CP model and impose a total order
over all operations (the I for Isolation property of ACID)?
No, this would be overkill.
Different operations have different requirements, and even for a
CAP-sensitive invariant, not all executions need to be synchronised.

%
For instance, since the \code{get-*} operations are read-only, they do not
change the truth of the \code{process-prescription} precondition.
Furthermore, even though \code{update-prescription-medication} changes
the \code{count} of a medication, it only increases it, which cannot
negate the precondition.
In other words, the precondition of \code{process-prescription} is
stable under concurrent \code{get-*} or
\code{update-prescription-medication}, and it is safe to let them
run concurrently.

When a precondition is unstable, the developer has
exactly two alternatives:
\begin{inparaenum}[\it (i)]
 \item
 Either to forbid concurrency, in order to avoid negating the
 precondition check; the update runs in CP mode, at the expense of
 availability; 
 or
 \item
 to remain available, but accept that the invariant might be violated (in
 which case it is not a real invariant!).
\end{inparaenum}
%
If the FMKe developer chooses the first option, she instructs the
database to forbid two \code{process-prescription} operations concerning
the same prescription from running concurrently; then, a user will not
be able to get her medication when the network is partitioned.
The second option is to downgrade the invariant to a best-effort
objective, or even to remove the check altogether, and risk delivering a
medication in duplicate.
This is a design decision: it's a trade-off between the availability
\emph{of this particular operation} and the value \emph{of this
  particular invariant.}
In fact, for the designers of the FMK production system, availability
was the top design objective, and they chose the second option,
accepting a non-zero probability of delivering medication in duplicate.

If the developer wishes to make the opposite decision, and enforce a
CAP-sensitive invariant, what are her choices?
To remain as available as possible, we wish to synchronise \emph{only
  when strictly necessary}.
In this example, we would forbid running two concurrent
\code{process-prescription} relating to the same prescription, but allow
it for different prescriptions.
We would also let \code{process-prescription} run concurrently
with \code{get-*} or \code{update-prescription-medication}.

\subsection{Verifying general CAP-sensitive invariants}
\label{sec:CISE-tool}


We now understand why certain updates need to synchronise, and others
not.
But this is getting complicated.
How is a developer to get it right?
With too little synchronisation, invariants can be violated; with too
much, availability and performance suffer.
\citet{syn:1749} show that an ad-hoc approach is error prone.

%

The bad news is that, outside of the strongest CP models, avoiding
unstable invariants is not transparent and requires knowledge of the
application.
The good news is that we have developed tools to automate this analysis,
and ensure that there are no mistakes --- to verify statically, i.e., at
design time, that your invariants are verified, even though most
operations remain available.
No guesswork!


Consider our CISE tool \cite{syn:app:sh179, app:sh183}.
Given the application specification (expressed in first-order logic),
CISE checks the following conditions:
\begin{inparaenum}[\it (i)]
\item 
  operations are Correct Individually (see Section~\ref{sec:appl-invar});
\item
  concurrent updates converge (see Section~\ref{sec:data-model:-crdts}),
and
\item
  every precondition is stable with respect to
  concurrent updates.
\end{inparaenum}
If all three checks pass, this constitutes a formal proof that the
application invariant remains true at all times when the application
runs above a Causally Consistent database \cite{syn:app:sh179}.
Otherwise, the tool returns a counter-example, which the developer can
use to diagnose the cause of the problem.
The fix can either be to change the application semantics in order to
remain AP, or to synchronise the two updates (switching to CP, but only
when strictly necessary).

Checking the FMKe application runs like this.
The invariant to verify is that a medication is not delivered more times
than prescribed.
First, the tool verifies that, for every FMKe operation in isolation,
with any legal parameter value, its precondition implies the invariant.
This check passes, because \code{update-prescription-medication}
can only increase \code{count}
and because \code{process-prescription} checks the remaining \code{count} of
every medication, and decreases that \code{count} by what is delivered.

Second, it verifies that replicas will converge, by comparing that all
pairs of concurrent operations (with any parameter), yield the the same
database state when run in opposite orders.
This check passes, for the following reasons.
The \code{get-*} operations have no side effects, therefore they commute
with all operations.
The \code{create-prescription} is necessarily causally before any other
operation on the same prescription, hence not concurrent with it.
The \code{update-prescription-medication} and
\code{process-prescription} operations operate on CRDTs, which converge
by construction.

For precondition stability, the tool checks that no update, with any
argument, ever negates the precondition check of a concurrent update.
This check fails, returning the following counter-example.
It starts with a prescription containing a medication count of one, and
performing \code{process-prescription} twice concurrently.
The precondition check is not stable since one tests \code{count} to be
1, and the other changes it to 0.
This shows that, to maintain the invariant, \code{process-prescription}
must synchronise with other \code{process-prescription}s of
the same prescription.
If we add this synchronisation to the application, we can run the tool
again; this time the verification succeeds.
Alternatively (following the FMK design explained in
Section~\ref{sec:total-order}), we can remove the ``no duplicates''
invariant; this also causes verification to succeed.

In order to support the CISE analysis, Antidote will run specific
transactions in a CP mode that upholds both TCC and the ACID properties.

\section{Conclusion}
\label{sec:concl-future-work}


Developing correct and highly-scalable applications is a challenging
task.
Instead of shoe-horning an application to a rigidly-defined consistency
model, we advocate a new {Just-Right Consistency} approach, focusing on
\emph{maintaining the application invariants} that are already present in
a sequential environment.

Based on an appropriate data model, CRDTs, we showed which invariant
patterns are AP-compatible, and how they can be guaranteed transparently
in an AP system.
Accordingly, we recommend Transactional Causal Consistency as the
default consistency model.
Our Antidote open-source, CRDT-based database is the first one to fully
implement TCC\@.

The remaining patterns are sensitive to the CAP gap between safety and
availability.
The Bounded Counter (one of the data types supported by Antidote)
constitutes a pre-packaged solution to a common case, encapsulating the
necessary synchronisation and minimising its impact.
For the general CAP-sensitive case, the CISE logic and tools verifies
whether an application has sufficient synchronisation, and if not, helps
identify the offending operations.
This enables tailoring synchronisation precisely to the application
requirements.

\section*{Acknowldegements}

The Just-Right Consistency concept derives from previous work by
\citet{rep:syn:sh167} and has benefited from discussion with
\begin{inparablank}
\item Masoud Saieda Ardekani
  and
\item Alexey Gotsman.
\end{inparablank}
The Bounded Counter concept is due to \citet{extending-numeric-2015-09}.
The CISE logic is due to \citet{syn:app:sh179}; the CISE tool was
conceived and implemented by Mahsa Najafzadeh \cite{app:sh183}.
FMKe was designed based on discussion with Kresten Krab Thorup.

Thanks to the whole Antidote team, who made this work possible:
\begin{inparablank}
\item Deepthi Akkoorath,
\item Valter Balegas,
\item Manuel Bravo,
\item Tyler Crain,
\item Viktória Fördős,
\item Michał Jabczyński,
\item Zhongmiao Li,
\item Ali Shoker,
\item Gonçalo Tomás,
\item Alejandro Tomsic,
and
\item Peter Zeller.
\end{inparablank}

{This research is supported in part by European projects
   \begin{inparablank}
     \item
     \href{http://syncfree.lip6.fr/}{SyncFree (FP7 609\,551)}%
, and
   \item%
     \hhref{LightKone (H2020 732\,505)}{https://lightkone.eu/}%
   \end{inparablank}
} 
\small
\bibliographystyle{abbrvnat-no-url}
\bibliography{predef,Y1-dissemination,Y2-dissemination,Y3-dissemination,shapiro-bib-ext,shapiro-bib-perso,local}

\begin{thebibliography}{27}
\providecommand{\natexlab}[1]{#1}
\providecommand{\url}[1]{\texttt{#1}}
\expandafter\ifx\csname urlstyle\endcsname\relax
  \providecommand{\doi}[1]{doi: #1}\else
  \providecommand{\doi}{doi: \begingroup \urlstyle{rm}\Url}\fi

\bibitem[Abadi(2012)]{rep:1691}
D.~J. Abadi.
\newblock Consistency tradeoffs in modern distributed database system design:
  {CAP} is only part of the story.
\newblock \emph{IEEE Computer}, 45\penalty0 (2):\penalty0 37--42, Feb. 2012.

\bibitem[Ahamad et~al.(1995)Ahamad, Neiger, Burns, Kohli, and
  Hutto]{syn:mat:1025}
M.~Ahamad, G.~Neiger, J.~E. Burns, et~al.
\newblock Causal memory: definitions, implementation, and programming.
\newblock \emph{Distributed Computing}, 9\penalty0 (1):\penalty0 37--49, Mar.
  1995.

\bibitem[Akkoorath et~al.(2016)Akkoorath, Tomsic, Bravo, Li, Crain, Bieniusa,
  Pregui{\c c}a, and Shapiro]{rep:pro:sh182}
D.~D. Akkoorath, A.~Z. Tomsic, M.~Bravo, et~al.
\newblock {C}ure: Strong semantics meets high availability and low latency.
\newblock In \emph{Int.\ Conf.\ on Distributed Comp.\ Sys. (ICDCS)}, pp.
  405--414, Nara, Japan, June 2016.

\bibitem[Attiya et~al.(2017)Attiya, Ellen, and Morrison]{syn:rep:1738}
H.~Attiya, F.~Ellen, and A.~Morrison.
\newblock Limitations of highly-available eventually-consistent data stores.
\newblock \emph{IEEE Trans.\ on Parallel and Dist.\ Sys. (TPDS)}, 28\penalty0
  (1):\penalty0 141--155, Jan. 2017.

\bibitem[Bailis et~al.(2013)Bailis, Davidson, Fekete, Ghodsi, Hellerstein, and
  Stoica]{db:syn:1751}
P.~Bailis, A.~Davidson, A.~Fekete, et~al.
\newblock Highly available transactions: Virtues and limitations.
\newblock \emph{Proc. {VLDB} {E}ndow.}, 7\penalty0 (3):\penalty0 181--192, Nov.
  2013.

\bibitem[Bailis et~al.(2015)Bailis, Fekete, Franklin, Ghodsi, Hellerstein, and
  Stoica]{syn:1749}
P.~Bailis, A.~Fekete, M.~J. Franklin, et~al.
\newblock Feral concurrency control: An empirical investigation of modern
  application integrity.
\newblock In \emph{Int.\ Conf.\ on the Mgt.\ of Data (SIGMOD)}, pp. 1327--1342,
  Melbourne, Victoria, Australia, 2015.

\bibitem[Balegas et~al.(2015{\natexlab{a}})Balegas, Pregui{\c c}a, Rodrigues,
  Duarte, Ferreira, Najafzadeh, and Shapiro]{rep:syn:sh167}
V.~Balegas, N.~Pregui{\c c}a, R.~Rodrigues, et~al.
\newblock Putting consistency back into eventual consistency.
\newblock In \emph{Euro.\ Conf.\ on Comp.\ Sys.\ (EuroSys)}, pp. 6:1--6:16,
  Bordeaux, France, Apr. 2015{\natexlab{a}}.

\bibitem[Balegas et~al.(2015{\natexlab{b}})Balegas, Serra, Duarte, Ferreira,
  Shapiro, Rodrigues, and Pregui{\c c}a]{extending-numeric-2015-09}
V.~Balegas, D.~Serra, S.~Duarte, et~al.
\newblock Extending eventually consistent cloud databases for enforcing numeric
  invariants.
\newblock In \emph{Symp.\ on Reliable Dist.\ Sys.\ (SRDS)}, pp. 31--36,
  Montr{\'e}al, Canada, Sept. 2015{\natexlab{b}}.
\newblock Not open access.

\bibitem[Barbar{\'a}-Mill{\'a} and Garcia-Molina(1994)]{syn:1735}
D.~Barbar{\'a}-Mill{\'a} and H.~Garcia-Molina.
\newblock The demarcation protocol: A technique for maintaining constraints in
  distributed database systems.
\newblock \emph{The {VLDB} Journal, The Int.\ J.\ on Very Large Data Bases},
  3\penalty0 (3):\penalty0 325--353, July 1994.

\bibitem[{Basho, Inc.}(2016)]{riak}
{Basho, Inc.}
\newblock {R}iak {KV}: Distributed {NoSQL} database.
\newblock Website \url{http://basho.com/products/riak-kv/}, 2016.
\newblock Accessed 4 June 2016.

\bibitem[Corbett et~al.(2012)Corbett, Dean, Epstein, Fikes, Frost, Furman,
  Ghemawat, Gubarev, Heiser, Hochschild, Hsieh, Kanthak, Kogan, Li, Lloyd,
  Melnik, Mwaura, Nagle, Quinlan, Rao, Rolig, Saito, Szymaniak, Taylor, Wang,
  and Woodford]{rep:pan:1693}
J.~C. Corbett, J.~Dean, M.~Epstein, et~al.
\newblock {S}panner: {G}oogle's globally-distributed database.
\newblock In \emph{Symp.\ on Op.\ Sys.\ Design and Implementation (OSDI)}, pp.
  251--264, Hollywood, CA, USA, Oct. 2012.

\bibitem[DeCandia et~al.(2007)DeCandia, Hastorun, Jampani, Kakulapati,
  Lakshman, Pilchin, Sivasubramanian, Vosshall, and Vogels]{app:rep:optim:1606}
G.~DeCandia, D.~Hastorun, M.~Jampani, et~al.
\newblock {D}ynamo: {A}mazon's highly available key-value store.
\newblock In \emph{Symp.\ on Op.\ Sys.\ Principles (SOSP)}, volume~41 of
  \emph{Operating Systems Review}, pp. 205--220, Stevenson, Washington, USA,
  Oct. 2007.

\bibitem[Du et~al.(2014)Du, Iorgulescu, Roy, and Zwaenepoel]{db:syn:1752}
J.~Du, C.~Iorgulescu, A.~Roy, et~al.
\newblock Gentle{R}ain: Cheap and scalable causal consistency with physical
  clocks.
\newblock In \emph{Symp.\ on Cloud Computing}, pp. 4:1--4:13, Seattle, WA, USA,
  Nov. 2014.

\bibitem[Gilbert and Lynch(2002)]{rep:pan:1628}
S.~Gilbert and N.~Lynch.
\newblock Brewer's conjecture and the feasibility of consistent, available,
  partition-tolerant web services.
\newblock \emph{SIGACT News}, 33\penalty0 (2):\penalty0 51--59, 2002.
\newblock ISSN 0163-5700.

\bibitem[Gotsman et~al.(2016)Gotsman, Yang, Ferreira, Najafzadeh, and
  Shapiro]{syn:app:sh179}
A.~Gotsman, H.~Yang, C.~Ferreira, et~al.
\newblock {'C}ause {I}'m {S}trong {E}nough: Reasoning about consistency choices
  in distributed systems.
\newblock In \emph{Symp.\ on Principles of Prog.\ Lang.\ (POPL)}, pp. 371--384,
  St.~Petersburg, FL, USA, 2016.

\bibitem[Lakshman and Malik(2010)]{db:rep:1768}
A.~Lakshman and P.~Malik.
\newblock {C}assandra: A decentralized structured storage system.
\newblock \emph{Operating Systems Review}, 44\penalty0 (2):\penalty0 35--40,
  Apr. 2010.
\newblock W. on Large-Scale Dist.\ Sys.\ and Middleware (LADIS) 2009.

\bibitem[Lamport(1978)]{con:rep:615}
L.~Lamport.
\newblock Time, clocks, and the ordering of events in a distributed system.
\newblock \emph{Communications of the {ACM}}, 21\penalty0 (7):\penalty0
  558--565, July 1978.

\bibitem[Li et~al.(2012)Li, Porto, Clement, Gehrke, Pregui{\c c}a, and
  Rodrigues]{rep:syn:1690}
C.~Li, D.~Porto, A.~Clement, et~al.
\newblock Making geo-replicated systems fast as possible, consistent when
  necessary.
\newblock In \emph{Symp.\ on Op.\ Sys.\ Design and Implementation (OSDI)}, pp.
  265--278, Hollywood, CA, USA, Oct. 2012.

\bibitem[Lloyd et~al.(2011)Lloyd, Freedman, Kaminsky, and
  Andersen]{rep:syn:1662}
W.~Lloyd, M.~J. Freedman, M.~Kaminsky, et~al.
\newblock Don't settle for eventual: scalable causal consistency for wide-area
  storage with {COPS}.
\newblock In \emph{Symp.\ on Op.\ Sys.\ Principles (SOSP)}, pp. 401--416,
  Cascais, Portugal, Oct. 2011.

\bibitem[Lloyd et~al.(2013)Lloyd, Freedman, Kaminsky, and
  Andersen]{syn:rep:1708}
W.~Lloyd, M.~J. Freedman, M.~Kaminsky, et~al.
\newblock Stronger semantics for low-latency geo-replicated storage.
\newblock In \emph{Networked Sys.\ Design and Implem.\ (NSDI)}, pp. 313--328,
  Lombard, IL, USA, Apr. 2013.

\bibitem[Najafzadeh et~al.(2016)Najafzadeh, Gotsman, Yang, Ferreira, and
  Shapiro]{app:sh183}
M.~Najafzadeh, A.~Gotsman, H.~Yang, et~al.
\newblock The {CISE} tool: Proving weakly-consistent applications correct.
\newblock In \emph{W.\ on Principles and Practice of Consistency for Distr.\
  Data (PaPoC)}, EuroSys 2016 workshops, London, UK, Apr. 2016.

\bibitem[O'Neil(1986)]{db:syn:1689}
P.~E. O'Neil.
\newblock The escrow transactional method.
\newblock \emph{Trans.\ on Database Systems}, 11\penalty0 (4):\penalty0
  405--430, Dec. 1986.
\newblock ISSN 0362-5915.

\bibitem[Shapiro et~al.(2011)Shapiro, Pregui{\c c}a, Baquero, and
  Zawirski]{syn:rep:sh143}
M.~Shapiro, N.~Pregui{\c c}a, C.~Baquero, et~al.
\newblock Conflict-free replicated data types.
\newblock In \emph{Int.\ Symp.\ on Stabilization, Safety, and Security of
  Dist.\ Sys.\ (SSS)}, volume 6976 of \emph{Lecture Notes in Comp.\ Sc.}, pp.
  386--400, Grenoble, France, Oct. 2011.

\bibitem[Shapiro et~al.(2016)Shapiro, Saeida~Ardekani, and
  Petri]{syn:formel:sh190}
M.~Shapiro, M.~Saeida~Ardekani, and G.~Petri.
\newblock Consistency in {3D}.
\newblock In \emph{Int.\ Conf.\ on Concurrency Theory (CONCUR)}, volume~59 of
  \emph{Leibniz Int.\ Proc.\ in Informatics (LIPICS)}, pp. 3:1--3:14,
  Qu{\'e}bec, Qu{\'e}bec, Canada, Aug. 2016.

\bibitem[{The SyncFree Consortium}()]{antidote-website}
{The SyncFree Consortium}.
\newblock {A}ntidote{DB}: A planet-scale, available, transactional database
  with strong semantics.
\newblock Website \url{http://antidoteDB.eu/}.

\bibitem[Tom{\'a}s et~al.(2017)Tom{\'a}s, Zeller, Balegas, Akkoorath, Bieniusa,
  Leit{\~a}o, and Pregui{\c c}a]{app:rep:1774}
G.~Tom{\'a}s, P.~Zeller, V.~Balegas, et~al.
\newblock {FMKe}: a real-world benchmark for key-value data stores.
\newblock In \emph{W.\ on Principles and Practice of Consistency for Distr.\
  Data (PaPoC)}, Belgrade, Serbia, Apr. 2017.

\bibitem[Zawirski et~al.(2015)Zawirski, Pregui{\c c}a, Duarte, Bieniusa,
  Balegas, and Shapiro]{rep:pan:sh177}
M.~Zawirski, N.~Pregui{\c c}a, S.~Duarte, et~al.
\newblock Write fast, read in the past: Causal consistency for client-side
  applications.
\newblock In \emph{Int.\ Conf.\ on Middleware (MIDDLEWARE)}, pp. 75--87,
  Vancouver, BC, Canada, Dec. 2015.

\end{thebibliography}

\end{document}